\documentclass[prl,twocolumn,superscriptaddress,showpacs]{revtex4}

\usepackage{graphics}           
\usepackage{bm}                 
\usepackage{amsmath}            
\usepackage{verbatim}           
\usepackage{color}
\usepackage{amsfonts,amssymb,latexsym,amscd}
\usepackage{amsthm}             
\theoremstyle{plain}            

\begin{document}

\title{Maximum Power Efficiency and Criticality in Random Boolean Networks}


\author{Hilary A. Carteret}
\affiliation{Institute for Biocomplexity and Informatics, Biosciences,
University of Calgary, Calgary, Alberta, T2N 1N4, Canada.}
\email{hcartere@qis.ucalgary.ca}

\author{Kelly John Rose}
\affiliation{Institute for Biocomplexity and Informatics, Biosciences,
University of Calgary, Calgary, Alberta, T2N 1N4, Canada.}
\affiliation{Department of Mathematics and Statistics, \\
University of Calgary, Calgary, Alberta, T2N 1N4, Canada.}
\email{kjrose@gmail.com}

\author{Stuart A. Kauffman}
\affiliation{Institute for Biocomplexity and Informatics, Biosciences,
University of Calgary, Calgary, Alberta, T2N 1N4, Canada.}
\email{skauffman@ucalgary.ca}

\date{October 3, 2008}

\begin{abstract}
Random Boolean networks are models of disordered causal systems that 
can occur in cells and the biosphere.  These are open thermodynamic 
systems exhibiting a flow of energy that is dissipated at a finite 
rate.  Life does work to acquire more energy, then uses the available 
energy it has gained to perform more work.  
It is plausible that natural selection has optimized many biological 
systems for \emph{power} efficiency: useful power generated per unit 
fuel.  In this letter we begin to investigate these questions for random 
Boolean networks using Landauer's erasure principle, which defines a 
minimum entropy cost for bit erasure.   
We show that critical Boolean networks maximize available power 
efficiency, which requires that the system have a finite displacement 
from equilibrium.  
Our initial results may extend to more realistic models for cells and 
ecosystems. 
\end{abstract}



\maketitle


\paragraph{Introduction:}

Random Boolean networks (RBNs) are a powerful class of models for 
complex causal systems \cite{epigen,origins}.  
However, a thermodynamics for such networks has not yet been developed. 
In this paper we define a minimum rate of energy flow for general RBNs 
and show that dynamically critical RBNs maximize \emph{power} efficiency. 

A Boolean network (BN) consists of $n$ nodes that output one bit each 
per time-step.  Each node $i$ receives inputs from $k_i$ nodes and uses 
a Boolean function $f_i$ that can be defined thus: 
\begin{equation}
 f_i:\{0,1\}^{k_i} \mapsto \{0,1\}
\end{equation}
to compute its output.  A BN model is logically equivalent to the system 
it is simulating.  Random BNs are used to make statistical statements 
about systems whose logical structure is unknown (or incompletely known). 
The number of connections between the nodes of an RBN can come from any 
desired distribution.  The Boolean functions are assigned to the nodes 
randomly from a uniform distribution over the possible Boolean functions, 
and all the nodes' outputs update synchronously. 
A state of a Boolean network is the current outputs of the $n$ nodes; 
the system traces out a trajectory of state transitions until it 
reaches a state-cycle attractor. 

There exist a variety of metrics which can be used to characterize 
the behavior of Boolean nets.  One of these is the Hamming metric, 
which is the number of bits by which two states differ.  
For any of these metrics, there exists an ordered regime, where 
nearby states lie on trajectories that converge on average in state 
space, and a chaotic regime where nearby states have divergent 
trajectories \cite{origins}.  
These two regimes are separated by a critical surface (i.e.~a phase 
transition) on which such trajectories stay the same average Hamming 
distance apart.  (See \cite{macrophage} for an example of this effect 
in another metric.)

One way to derive thermodynamics starts from the definition of the 
Carnot cycle, which gives a criterion for maximizing \emph{energy} 
efficiency: the most energy efficient machine for performing work 
is a reversible one, i.e.~that a machine must be at least as 
energy-inefficient as it is irreversible \cite{smith}.  
We show that by using a minimal notion of reversibility and 
hence an entropy production rate for RBNs, it is possible to 
define such a thermodynamics in terms of a \emph{specific dissipation rate} 
\cite{onsager31I,onsager31II}.

In the next two sections we outline some of the concepts we use in the 
paper and develop them into formal definitions in the following section.  
Using Landauer's erasure principle \cite{landauer61} we derive expressions 
for the minimum (intrinsic) entropy production rate and hence the maximum 
possible \emph{power} efficiency for RBNs and demonstrate that the power 
efficiency of RBNs is maximized when the network is critical.
We finish by proposing a ``maximum power principle'' for RBNs and 
discuss some possible further developments of this work.

\paragraph{Logically reversible computation and Landauer's principle}

A reversible Turing machine \cite{lecerf,bennett82} is similar to a 
conventional Turing machine (TM); both of them have a \emph{tape} that 
consists of a string of ``cells'', each of which can contain only one bit.  
The tape is used as a memory and initially may contain input data 
(if used) that is accessed by a read-write \emph{head}, which can move 
left or right and passes information to and from the processor, which 
consists of a \emph{transition function} (or \emph{table}) that is a 
finite set of instructions (the program) together with the 
\emph{state register} that stores the state of the processor, thus 
enabling the TM to keep track of where it is in its table.  
Unlike a conventional TM, a reversible TM is also required to be able to 
run backwards, i.e., at every step of the computation, the immediately 
preceding state must be uniquely defined.  
If we think of the TM as tracing out paths in a space of possible states, 
this requirement means that whenever two paths in this state-space merge, 
the TM must somehow keep track of which path it followed to reach that 
point.  
This information is called the \emph{history} of the computation and can 
be thought of as being written on a second tape.  (One tape is sufficient, 
but two are often used for pedagogical purposes.)

The only logically irreversible step in a computation is the erasure of 
information.  Landauer's erasure principle \cite{landauer61} is a corollary 
of Boltzmann's definition for the entropy, $S=k_{\text{B}}\ln \Omega,$ 
and states that a logically irreversible erasure is also thermodynamically 
irreversible.  Thus the irreversible erasure of a completely random, 
unknown, binary bit of information must generate an amount of entropy 
not less than $k_{\text{B}}\ln 2.$ 

The definition of an irreversible erasure for a logical bit is subtle, 
not least because there may be multiple, redundant copies of each bit 
in the machine's memory.
In order for a bit to survive a computational step, we must be able to 
find a copy of it after the step in a specified location that only 
depends on the program being executed, but not on the values of the 
other variables.  If the information in the bit ends up in different 
places depending on some other variables, it cannot be retrieved in this 
way \cite{LiVi96}.
The ability to retrieve the logical bit at a later time is essential; 
even with a complete description of a given TM, this entire-machine 
erasure rate is an uncomputable function, even for computations that 
halt \cite{zurek89}.  

After the machine has completed its calculation and copied its result, it 
must somehow return to its initial state in order to close the work-cycle.  
However, it cannot simply reset its memory, as this would require logically 
irreversible erasures (i.e., dissipation).
The only way the machine can clean up its memory without erasing anything 
is to ``uncompute'' the calculation it has just performed and run 
backwards to its initial state.

\paragraph{A minimal thermodynamics for RBNs:}

The original motivation for reversible computation theory was to 
determine if computation required a non-zero rate of energy dissipation. 
The fact that it doesn't led Bennett to compare a reversible computation 
to a Carnot cycle \cite{bennett82,feynman}. 

Bennett's results were obtained for Turing machines;
strictly speaking, a Boolean network is not a Turing machine.  
However, a BN can be described as a network of finite TM nodes operating 
in parallel with each node computing the value of its Boolean logic 
function. 
The nodes can output at most one bit at each computing cycle, though 
they are permitted to make copies of that bit and distribute them along 
the edges of their network di-graph to some number of other nodes. 
The TM at each node has a $k$-bit memory (part of its tape) 
for the values on the input edges to that node,
and it only needs a finite size working space on its tape 
(as any Boolean function on $k$ inputs can always be implemented by 
something no bigger than a finite-size look-up table, thus needs only a 
finite memory).
Thus we need only consider a finite-size, $k$-bit input TM at each node; 
the asymptotic limit in our analysis is the limit when the number of 
nodes tends to infinity, not when $k$ tends to infinity. 
This is fortunate, because if each node was a full-size TM, the erasure 
rate at each node would also be an uncomputable function (by Zurek's 
argument in \cite{zurek89}) as well as the erasure rate for the infinite 
BN as a whole.

For a Boolean function with a probability $p$ of outputting a $1$ 
(called the ``bias'' of the function) erasure produces a change 
in entropy $\Delta S$ of at least $-k_{\text{B}}\mathsf{S}(p),$ where 
$\mathsf{S}(p)$ is the Shannon entropy \cite{shannon} of the bit.
As long as the values of the nodes continue to change, the BN has not 
halted, even after it has entered a limit cycle. 
The BN will continue to erase bits every time it passes though a point 
in state space where a trajectory joins that limit cycle, because 
the system cannot remember what its precursor state was without 
storing at least one history bit \cite{bennett73}.  Thus these history 
bits must be either stored or erased whenever a tributary trajectory 
joins the limit cycle.  Landauer's principle dictates that erasing 
these history bits must cause dissipation.

Furthermore, reversible TMs can only achieve computation without 
dissipation if proceeding at zero speed \cite{bennett82,feynman}.
Fortunately, Bennett extended his rules to finite-speed computation 
\cite{bennett73,bennett82,CHBnotes88,feynman,LiVi96}. 
As soon as the computation leaves this adiabatic limit, the laws 
governing the entropy generation rate change: 
computation at finite speed requires dissipation, just as the ideal 
energy efficiency for any work-cycle can only be achieved in the 
adiabatic limit, when it is performed quasi-statically.

If our BN is to run at a non-zero speed for a useful length of time, 
we must supply power to drive it.  
For example, consider a BN implemented as a set of coupled chemical 
reactions: the concentrations of the reagents will only fluctuate 
around their equilibrium values unless active steps are taken to drive 
the system away from equilibrium.
For a reversible TM to be $r$ times more likely to move forwards as 
backwards requires $k_{\text{B}} T \ln r$ energy to be dissipated per 
computational step \cite{bennett73,feynman}.
The need to drive the BN with some (generalized) force is a general 
fact about reversible computation at finite speed and is  
completely independent of its implementation; it holds whether the 
BN is a genetic network or something else.  
Only the dimensionless expression for $r$ will depend on these details.
This implies an energy cost for copying; while logically reversible, 
it can only be done for free if performed infinitely slowly 
\cite{bennett82,feynman}.
The faster the copies are produced, the more energy they will cost, 
even if the register into which each copy is written was blank.

For processes that must occur in a finite time, a more natural 
(and useful) measure is the power efficiency, which is given by the 
\emph{Gouy-Stodola theorem} \cite{bejanRev}:
\begin{equation}\label{GSthm}
  \frac{dW_{\text{rev}}}{dt}-\frac{dW_{\text{use}}}{dt} 
       = T\frac{dS}{dt}
\end{equation}
where $T$ is the temperature of the environment into which the entropy 
$S$ is released.  
The Gouy-Stodola efficiency is maximized when the rate of entropy 
production per unit of power supplied to the system is minimized 
\cite{bejanRev}.  

We seek expressions for expected (mean-field) values for an otherwise 
unknown RBN drawn at random from some ensemble defined by some 
macroscopic parameters, such as the bias, $p.$ 
We therefore have only partial information about this RBN.
Since BNs evolve in a discrete time steps, we use a finite-difference 
version of this equation.  
and a mean-field approximation to the discrete dynamics to 
calculate the \emph{expected entropy production rate}.  
As this method finds an average value per node, we will 
write the mean $k$ instead of $k_i$ for the in-degree of a 
typical node $i.$  This approach should also allow discussion of the 
expected power requirements of the RBN under different initial conditions.

Reversible simulation of an irreversible TM incurs overheads in either 
time, space, or both \cite{bennett89,timespaceBTV}. 
Since we must supply power to the BN for each time step \cite{feynman}, 
any increase in the time taken will cost us, so we focus on 
time-parsimonious simulations.
A large space overhead \cite{timespaceBTV} raises the problem of where 
that information can be stored.  
If there is insufficient memory, then some bits will need to be erased 
\cite{LiVi96}.  Even so, additional memory can only delay the inevitable, 
after which the rate of copying information cannot exceed the erasure 
rate; writing a bit of information requires somewhere to write it. 
For each directed edge leaving a node, a copy is made of that node's 
bit onto that edge; thus for a node with out-degree $m,$ and since $m=k,$ 
$k$ edges must be prepared locally.
We also assume that every node has at least one incoming edge, without 
loss of generality.

In 2004, Shmulevich and Kauffman \cite{ShmuStu} defined a new measure of 
the average dynamical properties of BNs that they called 
the \emph{sensitivity}, $s$ of the Boolean function $f_i,$ which is the 
number of inputs to $f_i$ for which flipping that input bit alone changes 
the value of $f_i.$  Such input bits are said to be \emph{relevant} to 
that node.
The more sensitive a BN is, the better it remembers perturbations; thus 
the rate it erases the information that distinguishes one computational 
path from another is low.

Shmulevich and Kauffman showed \cite{ShmuStu} that for a Boolean function 
of bias $p,$ the probability that an input edge is relevant to any given 
node (its ``activity'') is $\alpha=2p(1-p).$  
Whether or not an input bit is locally erased at a particular node depends 
on its \emph{relevance} to that node.  This will typically depend on the 
values of other input edges.  However, the irrelevant bits may not have 
been erased from the network \emph{as a whole}.  
If an edge is \emph{irrelevant}, the local copy on that edge is lost, 
since it is not recoverable by any procedure that is independent of the 
state of the larger RBN, if at all.  
If all edge-copies of a bit are irrelevant under at least one set of 
values of the other edges (it need not be the same set) then that 
bit is irreversibly erased \cite{LiVi96}.

\paragraph{A lower bound for the expected entropy production rate per node:}

To make an BN as efficient as possible, we must minimize the number of 
irreversible erasures at each step, or equivalently, the number of 
copies used by each node per step.
We use Landauer's principle and the non-zero energy cost of copying 
a bit at a finite speed \cite{feynman} to find a lower bound for the 
entropy production rates for RBNs. 
Writing Eq.~\eqref{GSthm} in finite difference form gives
\begin{equation}\label{finiteGSthm}
  \frac{\Delta W_{\text{in}}}{\Delta t} = 
   \frac{T\Delta S}{\Delta t} + \frac{\Delta W_{\text{use}}}{\Delta t} 
\end{equation}
where $T \Delta S$ is the dissipated energy lost due to Landauer's  
principle and $\Delta W_{\text{use}}$ is the free energy available per 
step to drive the computation forward. 
We divide~\eqref{finiteGSthm} by the finite difference ratio 
$\Delta W_{\text{rev}}/\Delta t$ to obtain the corresponding power 
efficiency measure:
\begin{equation}\label{peffdef}
  E_{\text{P}}:=\frac{\Delta W_{\text{use}}}{\Delta W_{\text{rev}}}
    = \frac{\Delta W_{\text{rev}}-T\Delta S}{\Delta W_{\text{rev}}} 
    = 1-\frac{T \Delta S}{\Delta W_{\text{rev}}}.
\end{equation}

Let $A_{\text{in}}$ be the expected number of edges that need to be 
prepared for each node for each time-step.  Let $A_{\text{out}}$ be the 
expected number of bits of information erased per node per time step as 
a result of the computation. 
Then $\mathcal{D},$ the \emph{specific dissipation rate per node} is  
\begin{equation}
 \mathcal{D}:= 
   \left|(A_{\text{in}} - A_{\text{out}})/A_{\text{in}}\right|,
\end{equation}
which will give us $E_\text{P} = 1 - \mathcal{D}T.$  
(The modulus function is needed because for some ranges of the parameters 
there is a net loss of information at the node, whereas for others, the 
network is sending bit-copies into the node's equivalent finite TM that 
are not erased at this time-step since they are relevant to the function 
at that node.)

Shmulevich and Kauffman also demonstrated \cite{ShmuStu} that the 
sensitivity can be defined for every BN. 
They calculated this value to be $s=k\alpha$ and show that when $s=1$, 
the RBN is critical.  We use this to show that the dissipated power 
$\mathcal{D}$ is minimized for critical networks, and thus $E_\text{P}$ 
is maximized accordingly.
Since the expected activity of each edge is $\alpha$ we see that on 
average every node will have at least $k(1-\alpha)+1 = k-s+1$ copies 
which must be deleted in the process of computing the Boolean function 
for the associated node, since they are inactive at that node.
(The additional $+1$ is from the deletion of the output bit from the 
previous time-step.)
Since these input bits are irrelevant, their information content cannot 
be retrieved from the output of that node.
Since $A_{\text{in}}=k,$ 
this allows us to write down an expression for the expected rate of 
information loss per node:
\begin{equation}\label{Aouteq}
 A_{\text{out}} = k(1-\alpha)+1 = k-k\alpha+1 = 1+k-s.
\end{equation}
Using Eq.~\eqref{Aouteq}, it is possible to approximate the erasure rate 
$\mathcal{D}$ for a BN given any two of $p, s,$ or $k.$  
\begin{equation}
  \mathcal{D} = \left|(k - (1+k(1-\alpha)))/k\right| 
              = \left|\alpha(s-1)/s\right|,
\end{equation}
where $k=s/\alpha.$
Thus when $s=1$ (i.e. $k=\alpha^{-1}$) we have $\mathcal{D}=0$ at the 
critical line. 
Since $\mathcal{D}$ is the average erasure rate per node, the power 
efficiency $E_\text{P}$ is maximized at the dynamical critical line for RBNs. 
As long as every node has at least one incoming edge, the minimum lower 
bound for $\mathcal{D}=0$ occurs at critical values of $k.$

\paragraph{A ``maximum power principle'' and other questions:}

\begin{figure}
    \begin{minipage}{\columnwidth}
    \begin{center}
        \resizebox{0.8\columnwidth}{!}{\includegraphics{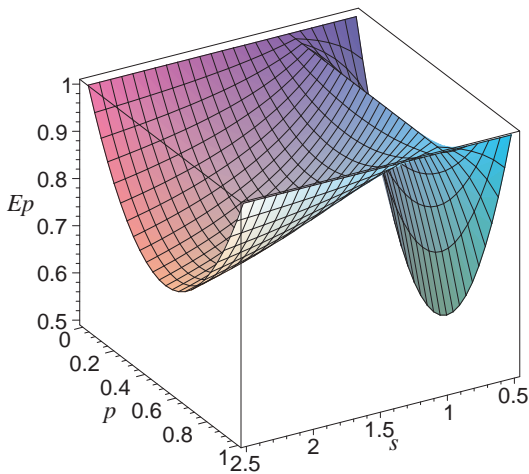}}
    \end{center}
    \end{minipage}
    \caption{Power efficiency $E_{\text{P}}$ at $T=1,$ as a function 
             of $s$ and $p,$ showing maximized power efficiency 
             in a ridge at criticality; 
             $E_{\text{P}}$ is maximized at the critical line of 
             Derrida and Pomeau \cite{derrida}.
             As $p \rightarrow 0$ or $1$ the input bits at each node have 
             a lower Shannon entropy, so less information is lost if 
             the output bit is subsequently erased.  Thus 
             $E_{\text{P}} \rightarrow 1$ when $p \rightarrow 0$ or $1.$   
             }\label{spplot}
\end{figure}

Suppose we have an expression for the power available to drive the RBN, 
$\Delta W_{\text{in}}/\Delta t.$  
A reversible TM computing at a rate $r$ consumes energy 
$k_{\text{B}} T \ln r$ per time step.
The form of the rate function $r$ will depend on the implementation 
of the RBN.  
Thus we have an expression for the maximum possible ``metabolic rate'' 
per node for the RBN with bias $p,$
\begin{equation}
  r = \mathsf{S}(p) e^{-\mathcal{D}} e^{P_{\text{in}}\Delta t/k_{\text{B}}T}.
\end{equation}
The RBNs most closely approaching these upper bounds are dynamically 
critical; this suggests the existence of a ``maximum power principle'' 
\cite{odum,lotka22a} for RBNs.
Figure~\ref{spplot} shows how $E_{\text{P}}$ varies with $s$ and $p$ when 
$k$ is fixed; similar behavior occurs for other distributions of $k$. 

It should be noted that the distinction between the copy-cost and the 
available energy cost may not be sharp in some implementations.
In the molecular computer mentioned above, 
the driving power is supplied via the chemical potential terms in the 
Gibbs free energy. Maintaining the reactants in such non-equilibrium 
concentrations requires the larger system to do work.

\paragraph{Conclusion}

We have shown that critical RBNs maximize power efficiency; 
critical RBNs also maximize pairwise mutual information \cite{mutInfRBN}.
We conjecture a direct causal link between these three phenomena.  
Moreover Odum has proposed a maximum power principle for ecosystems 
\cite{odum} and Ulanowicz has argued that mature ecosystems 
maximize their mutual information content \cite{Uvariation87}. 
Furthermore, cells may be critical \cite{macrophage} and the 
rate of accumulation of biomass in an ecosystem may also be maximized 
at criticality \cite{microbe,jorgensen}.
In more detailed and realistic models that include more modes of 
energy expenditure, our lower bound should increase. 
Future work will look for direct relationships between criticality, 
mutual information and power efficiency in causal networks. 


\begin{acknowledgments}

\noindent
This work was supported by i{\small{CORE}}, NSERC and the 
University of Calgary.  The authors would also thank 
Mircea Andrecut, Ilya Shmulevich and Matti Nykter 
for some stimulating discussions.

\end{acknowledgments}





\end{document}